\documentclass[floatfix,aps,twocolumn,preprintnumbers,amsmath,amssymb]{revtex4}

\usepackage{amsfonts, amssymb, amsmath, dsfont}
\usepackage{graphicx}
\usepackage{float}
\usepackage{color}
\usepackage[utf8]{inputenc}
\usepackage[T1]{fontenc}

\newcommand{\bm}[1]{\boldsymbol{\mathbf{#1}}} 
\newcommand{\bra}{\langle}
\newcommand{\ket}{\rangle}

\newcommand{\de}{\operatorname{d}}
\newcommand{\re}{\operatorname{Re}}
\newcommand{\im}{\operatorname{Im}}

\newcommand{\eq}[1]{Eq.~\eqref{#1}}

\newcommand{\fig}[1]{Fig.~\ref{#1}}

\newcommand{\sect}[1]{Section~\ref{#1}}

\begin{document}

\title{Quantum dipole emitters in structured environments: A scattering approach (tutorial)}

\author{Dorian Bouchet}
\author{R\'emi Carminati}
\altaffiliation{remi.carminati@espci.fr}
\affiliation{ESPCI Paris, PSL University, CNRS, Institut Langevin, 1 rue Jussieu, F-75005, Paris, France}






\begin{abstract}
We provide a simple semi-classical formalism to describe the coupling between one or several quantum emitters and a structured environment. Describing the emitter by an electric polarizability, and the surrounding medium by a Green function, we show that an intuitive scattering picture allows one to derive a coupling equation from which the eigenfrequencies of the coupled system can be extracted. The model covers a variety of regimes observed in light-matter interaction, including weak and strong coupling, coherent collective interactions, and incoherent energy transfer. It provides a unified description of many processes, showing that different interaction regimes are actually rooted on the same ground. It can also serve as a basis for the development of more refined models in a full quantum electrodynamics framework. 
\end{abstract}



\maketitle

\section{Introduction}

Many aspects of light-matter interaction can be understood from the coupling between dipole emitters (or absorbers) and the electromagnetic field in a structured medium. Indeed, the basic processes in molecular spectroscopy, light scattering from small particles or atoms, fluorescence, nonlinear optics or cavity quantum electrodynamics (QED) are most of the time described based on electric (or magnetic) dipoles interacting with the electromagnetic field~\cite{cohen-tannoudji_atomphoton_2010,mandel_optical_1995,haroche_cavity_1992,haroche_exploring_2013}. With the advent of nanophotonics, structuring the environment at scales much smaller than the wavelength is used to modify and control the emission and absorption dynamics of quantum emitters (such as molecules or quantum dots). This has become an active area of research, with fundamental and applied perspectives \cite{novotny_principles_2012,koenderink_nanophotonics:_2015}.

Depending on the strength of the interaction, different regimes are observed. In the weak coupling regime, spontaneous emission can be either accelerated or inhibited, a phenomenon referred to as the Purcell effect \cite{purcell_spontaneous_1946}. When the emitter strongly couples to a specific mode of the electromagnetic field, two new hybridized eigenmodes (polaritons) are created, characterized by a frequency splitting or the appearance of Rabi oscillations in the time domain~\cite{allen_optical_1987,barnes_special_2018}. Initially the realm of cavity QED, changes in the spontaneous emission dynamics in the weak and strong coupling regimes has been demonstrated in nanophotonics using optical antennas \cite{novotny_antennas_2011},
microcavities \cite{weisbuch_observation_1992,kavokin_cavity_2003}, photonic crystal cavities \cite{yoshie_vacuum_2004}, or plasmonic cavities \cite{chikkaraddy_single-molecule_2016}. The mutual interaction between several emitters in the presence of an electromagnetic field also gives rise to different phenomena, from energy transfer between two molecules in weak coupling~\cite{medintz_fret_2013}, to coherent collective interactions leading to sub and superradiance~\cite{agarwal_quantum_1974,gross_superradiance:_1982}. Here as well, confining the electromagnetic field allows one to act on the coupling strength. For example, the range of energy transfer can be modified using surface plamons~\cite{bouchet_long-range_2016}, and collective interactions can be enhanced using photonic crystal cavities \cite{sipahigil_integrated_2016}.

In this tutorial, we propose a simple and unified approach to deal with the interaction between a quantum emitter and the electromagnetic field in a structured medium, and we show how the same starting point allows one to describe many different regimes and phenomena in light-matter interaction. The emitter is described by an electric polarizability and the field is described in terms of a Green function. Assuming an external excitation, we address the coupling as a semi-classical scattering process (by semi-classical we mean that the field is not explicitly quantized), and we derive a coupling equation from which the eigenfrequencies of the resulting eigenmodes can be deduced. By choosing the correct model for the Green function, which describes the response of the environment, the formalism naturally leads to a description of the weak and strong coupling regimes. The intuitive scattering approach is easily extended to the situation of two emitters coupled through a structured environment. Interestingly, beyond coherent mutual interactions leading to strong coupling, the model also includes a description of incoherent energy transfer between molecules in the weak coupling regime. Finally, we show how a generalization to a set of $N$ emitters provides an appealing coupled-dipole model to describe collective interactions. 

The tutorial is organized as follows. In \sect{sect_polar}, starting from the optical Bloch equations, we derive the polarizability model that allows us to describe either the full dynamics of a two-level atom or the excitation dynamics of a three-level molecule. In \sect{sect_Green}, we introduce the concept of Green function, which is a useful tool to describe the electrodynamic response of an arbitrary environment. In \sect{sect_one_emitter}, we derive the coupling equation that drives the dynamics of the coupled emitter-field system, based on an intuitive scattering approach. From this equation, we show how the weak and strong coupling regimes emerge. In \sect{sect_two_emitters}, we extend the scattering approach to the situation of two emitters coupled through a structured environment, focusing the analysis on the regimes of weak and strong dipole-dipole interaction. In the weak coupling regime, we show how irreversible energy transfer can be described using appropriate polarizability models. In \sect{sect_many_emitters}, we briefly discuss the generalization of the model to the collective interaction between $N$ identical emitters, with $N$ arbitrary large. Finally, \sect{sect_conclusion} summarizes the main conclusions.

\section{Polarizability of a dipole emitter}
 \label{sect_polar}

The electrodynamic response of a subwavelength resonant scatter can be described in the electric-dipole limit using a dynamic polarizability. The same description holds for an atom or a fluorescent molecule. The interaction between a two-level atom and a classical monochromatic electric field is a textbook problem, that is usually treated by solving the optical Bloch equations~\cite{allen_optical_1987,cohen-tannoudji_atomphoton_2010}. Here we use this framework to describe the excitation of a three-level system by a quasi-monochromatic electric field. The three-level model includes the two-level atom as a particular case. It also encompasses the main features needed to describe the excitation of a fluorescent molecule. 

\subsection{Three-level model \label{sect_three_level}}

\begin{figure}[htbp]
\centering
\includegraphics[width=3cm]{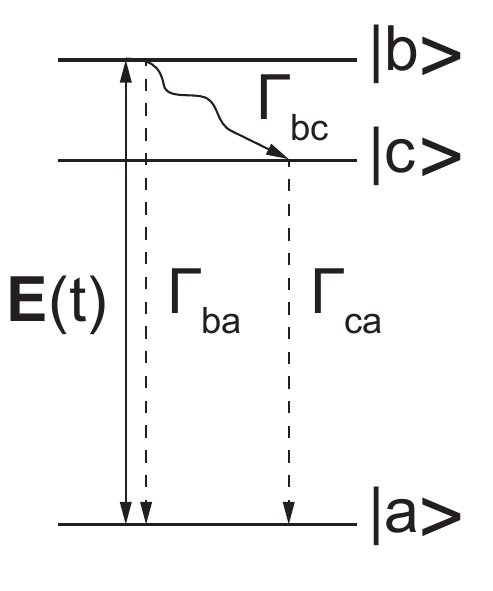}
\caption{Jablonski diagram of a three-level system. For $\Gamma_{bc}=0$ the system reduced to the model of a two-level atom. For $\Gamma_{bc} \gg \Gamma_{ba}$,
the three-level system is the simplest relevant model of a fluorescent molecule. In this case, $\Gamma_{bc}$ corresponds to a fast non-radiative decay towards state $|c\ket$ and $\Gamma_{ca}$ corresponds to the radiative transition.}
\label{fig_levels}
\end{figure}

We consider a three-level system characterized by three stationnary and non-degenerate eigenstates $| a \ket$, $| b \ket$ and $| c \ket$, as represented in~\fig{fig_levels}, with $\Gamma_{bc}$, $\Gamma_{ba}$ and $\Gamma_{ca}$ the spontaneous decay rates of each level. In practice, this three-level model can be used to describe a two-level atom (by taking $\Gamma_{bc}=0$), or a three-level system with a high decay rate towards the auxiliary level ($\Gamma_{bc} \gg \Gamma_{ba}$) that provides the simplest model of a fluorescent molecule.

The state of the system is conveniently described by a density operator $\hat{\rho}$. The diagonals elements of this operator, known as populations, give the probability for the system to be in one of its eigenstates. The off-diagonal elements, known as coherences, describe dynamic effects related to the coherent superpositions of eigenstates. They enter, as we shall see, the expression of the polarizability. The evolution of the density operator is driven by the Hamiltonian $\hat{H}$ according to \cite{cohen-tannoudji_quantum_1991}
\begin{equation}
\frac{\de \hat{\rho} }{\de t} =\frac{1}{i \hbar} [\hat{H},\hat{\rho}] \; .
\label{eq000}
\end{equation}
Using this equation is equivalent to using the Schr\"odinger equation for an arbitrary state $| \psi(t) \ket$ of the system, with the advantage of providing a straightforward description of mixed states. Since we are interested in the interaction between the three-level system and an external electric field, it is convenient to write $\hat{H}=\hat{H}_0+\hat{H}_1$ where $\hat{H}_0$ is the unperturbed Hamiltonian (describing the emitter in absence of electric field) and $\hat{H}_1$ is the interaction Hamiltonian (describing the coupling with the field). Following the procedure commonly used for two-level systems~\cite{mandel_optical_1995}, we can construct the unperturbed Hamiltonian for three-level systems, which is 
\begin{equation}
\hat{H}_0 = \hbar \omega_{ab} \hat{\sigma}_{ab}^+ \hat{\sigma}_{ab}^- +\hbar \omega_{ac} \hat{\sigma}_{ac}^+ \hat{\sigma}_{ac}^- \; ,
\end{equation}
where $\omega_{ij}$ is the Bohr frequency associated with the transition $ij$, and $\hat{\sigma}_{ij}^{+}=| j \ket \bra i | $ and $\hat{\sigma}_{ij}^-=| i \ket \bra j | $ are the atomic raising and lowering operators, respectively. In order to describe the excitation of the emitter, we use a semi-classical description and assume that it interacts with a classical quasi-monochromatic electric field tuned to the transition $ab$. In the electric-dipole approximation, we can express the interaction Hamiltonian as~\cite{cohen-tannoudji_atomphoton_2010,aspect_quantum_2010}
\begin{equation}
\hat{H}_1 =   - \bm{d}_{ab} \cdot \bm{E}(t) (\hat{\sigma}_{ab}^{+} + \hat{\sigma}_{ab}^-) \; ,
\label{eq00}
\end{equation}
where $\bm{E}(t)$ is the electric field at the position of the emitter and $\bm{d}_{ab}=\bra a |\hat{\bm{D}} | b \ket=\bra b |\hat{\bm{D}} | a \ket$ is the dipole matrix element (or transition dipole). At this stage we did not consider the effects of spontaneous emission and other interactions with the environment (such as collisions with other molecules in a gas, with phonons in a solid, or with internal degrees of freedom in the emitter itself). These processes, assumed to be independent of the external exciting field, affect the populations and the coherences, and need to be included in \eq{eq000}. This leads to the master equation
\begin{equation}
\frac{\de \hat{\rho} }{\de t} =\frac{1}{i \hbar} [\hat{H}_0+\hat{H}_1,\hat{\rho}] + \left \{  \frac{\de \hat{\rho} }{\de t} \right \}_{\mathrm{relax}}
\label{eq0}
\end{equation}
in which the last term accounts for the decay of populations and coherences due to spontaneous emission and dephasing processes (that contribute to the relaxation of the coherences). The form of the relaxation terms can be found by considering the three-level system in absence of an external driving field. In this case we know that the populations of states $| b \ket$ and $| c \ket$ decay spontaneously, with the rates indicated in~\fig{fig_levels}, allowing us to write $\de {\rho}_{bb}/\de t=- (\Gamma_{ba}+\Gamma_{bc}) \rho_{bb}$ and $\de {\rho}_{cc}/\de t=- \Gamma_{ca} \rho_{cc} + \Gamma_{bc} \rho_{bb}$. The coherences $\rho_{ba}$ and $\rho_{ab}=\rho_{ba}^*$, that will be needed to compute the polarizability in the next sections, also decay according to $\de \rho_{ba}/\de t=-  (\gamma_{ab}/2) \rho_{ba}$, with a damping rate $\gamma_{ab}$ satisfying 
$\gamma_{ab} \geq \Gamma_{ba}+\Gamma_{bc}$ (the equality holding only when pure dephasing processes, that do not change the energy states, can be neglected).
Note that formally, the last term in~\eq{eq0} can be represented by an operator $\hat{\mathcal{L}}_{d} (\hat{\rho})$ known as Lindblad superoperator, that is sometimes used to include explicitly
the relaxation terms in the master equation~\cite{carmichael_statistical_1999}.

\subsection{Optical Bloch equations}

Finding the solution to \eq{eq0} requires to solve a system of nine equations. For our purpose, we need to compute the excited-state populations $\rho_{aa}$, $\rho_{bb}$ and $\rho_{cc}$, as well as the coherences $\rho_{ab}$ and $\rho_{ba}$. The coherences will allow us to compute the expectation of the dipole moment operators associated to transition $ab$. Since the density operator is Hermitian and satisfies the condition $\rho_{aa}+\rho_{bb}+\rho_{cc}=1$, we can reduce the problem to a set of three equations. As we assume the external electric field to be quasi-resonant with transition $ab$, we can use the rotating wave approximation ($|\omega-\omega_{ab}|\ll \omega_{ab}$) and the slowly-varying envelope approximation ($\gamma_{ab}\ll \omega_{ab}$) \cite{mandel_optical_1995}. This leads to the optical Bloch equations \cite{allen_optical_1987,cohen-tannoudji_atomphoton_2010,aspect_quantum_2010}:
\begin{equation}
\frac{\de {\rho}_{bb}}{\de t}=- (\Gamma_{ba}+\Gamma_{bc}) \rho_{bb} + 2 \im \left[\rho_{ab} \,  \Omega^{(+)}(t) \right] \; ,
\label{eq_B1}
\end{equation}
\begin{equation}
\frac{\de {\rho}_{cc}}{\de t}=- \Gamma_{ca} \rho_{cc} + \Gamma_{bc} \rho_{bb} \; ,
\end{equation}
\begin{equation}
\frac{\de \rho_{ba}}{\de t}=-  \frac{\gamma_{ab}}{2} \rho_{ba} - i \omega_{ab} \rho_{ba} +  i (\rho_{bb}-\rho_{aa})\Omega^{(+)}(t) \; ,
\label{eq8}
\end{equation}
where we have introduced the time-dependent Rabi frequency $\Omega(t)=- [\bm{d}_{ab} \cdot \bm{E}(t)]/\hbar$, and its positive frequency component defined with the following convention:
\begin{equation}
\Omega^{(+)}(t)= \int_{0}^{+\infty} \Omega(\omega) e^{-i \omega t} \de \omega \; .
\label{eq16}
\end{equation}
The time-dependent Rabi frequency characterizes the coupling strength between the three-level emitter and the electric field. In the absence of an external field, $\Omega^{(+)}(t)=0$ and the system spontaneously decays towards its lower energy state $|a \ket$ due to the damping rates of populations and coherences. In contrast, in the presence of an electric field, the terms in $\Omega^{(+)}(t)$ couple the equations driving the populations and the coherences. In particular, \eq{eq_B1} shows that the evolution of the excited-state population depends on the phase difference between the coherences (related to the dipole moment operators) and the time-dependent Rabi frequency (related to the external field). In order to compute the polarizability associated with the transition $ab$, we need to solve the coupled Bloch equations and find the expression of the coherences $\rho_{ab}$ and $\rho_{ba}$.

\subsection{Polarizability}

Assuming an excitation by a stationnary external field, we focus on the steady-state behavior of the coupled emitter-field system. 
In this regime, the solution of the optical Bloch equations can be found analytically. Solving Eqs.~\eqref{eq_B1}-\eqref{eq8} in the frequency domain yields
\begin{equation}
 \rho_{ba}(\omega)=  - \frac{ \Omega^{(+)}(\omega)}{\omega_{ab}-\omega-i \gamma_{ab} /2}  \left( \frac{1}{1+s} \right) \; ,
 \label{eq9}
\end{equation}
where $s$ is the saturation parameter given by
\begin{equation}
s= \frac{2 (2 \Gamma_{ca}+\Gamma_{bc})}{\Gamma_{ca}(\Gamma_{ba}+\Gamma_{bc})}  \int_{-\infty} ^{+\infty} \im \left[ \frac{|\Omega^{(+)}(\omega')|^2  }{\omega_{ab}-\omega' -  i \gamma_{ab} /2} \right] \de \omega'  \; .
\end{equation}
Equation \eqref{eq9} can be used to compute the expectation value of the dipole moment operator defined as $\bm{d} = \mathrm{Tr}(\hat{\rho} \, \hat{\bm{D}})$. More precisely, in order to define a polarizability matching the classical convention for monochromatic fields with a time dependence $\exp(-i\omega t)$, we will need the positive frequency part of the expectation value that is given by $\bm{d}^{(+)}(\omega) = \rho_{ba}(\omega )\bm{d}_{ab} $. For weak exciting field we can neglect saturation effects ($s \ll 1$), and we obtain
\begin{equation}
\bm{d}^{(+)}(\omega) = \frac{1}{\hbar} \left( \frac{1}{\omega_{ab}-\omega-i \gamma_{ab}/2}   \right)  [\bm{d}_{ab} \cdot \bm{E}^{(+)}(\omega)] \, \bm{d}_{ab}  \; .
\end{equation}
By definition of the polarizability $\boldsymbol{\alpha}_{ab}(\omega)$, we also have
\begin{equation}
\bm{d}^{(+)}(\omega) = \boldsymbol{\alpha}_{ab}(\omega) \epsilon_0 \, \bm{E}^{(+)}(\omega)  \; .
\end{equation}
These two equations readily lead to the following expression of the polarizability characterizing the excitation of the three-level emitter:
\begin{equation}
\boldsymbol{\alpha}_{ab}(\omega)=\frac{3 \pi c^3}{\omega_{ab}^3}  \frac{\Gamma_{ba}^{sp}}{\omega_{ab}-\omega-i \gamma_{ab}/2}  \bm{u} \otimes \bm{u} \; .
\label{eq6}
\end{equation}
In this expression, we have introduced the unit vector $\bm{u}$ characterizing the orientation of the transition dipole, such that $\bm{d}_{ab} = d_{ab} \bm{u}$, and $\otimes$ denotes the tensor product. We have also introduced the spontaneous emission rate (or Einstein A coefficient)~\cite{mandel_optical_1995}
\begin{equation}
\Gamma_{ba}^{sp}=\frac{\omega_{ab}^3 d_{ab}^2}{3 \pi \epsilon_0 \hbar c^3} \; .
\end{equation}
Note that very often $\Gamma_{ba} \geq \Gamma_{ba}^{sp}$ since additional non-radiative processes can contribute to the decay of the excited-state $| b \ket$ towards the ground state $| a \ket$. From the expression of the polarizability $\boldsymbol{\alpha}_{ab}(\omega) = \alpha_{ab}(\omega)\bm{u} \otimes \bm{u} $, we can deduce the expressions of the extinction and scattering cross-sections 
$\sigma_{e}(\omega) =(\omega/c) \, \im \left[ \alpha_{ab}(\omega) \right] $ and $\sigma_{s}(\omega)= [\omega_0^4/(6 \pi c^4)] \,  \left \vert \alpha_{ab}(\omega) \right \vert^2 $~\cite{bohren_absorption_2008}. For quasi-resonant excitation ($\omega \simeq \omega_{ab}$), we have
\begin{equation}
\sigma_{e}(\omega) = \frac{3 \pi c^2}{2 \omega_{ab}^2} \, \frac{\gamma_{ab} \Gamma_{ba}^{sp}}{(\omega_{ab}-\omega)^2 + \gamma_{ab}^2/4}\; ,
\label{eq_extinction}
\end{equation}
\begin{equation}
\sigma_{s}(\omega) = \frac{3 \pi c^2}{2 \omega_{ab}^2} \, \frac{\left(\Gamma_{ba}^{sp}\right)^2}{(\omega_{ab}-\omega)^2 + \gamma_{ab}^2/4}\; .
\label{eq_scattering}
\end{equation}
Note that when the damping rate of the coherences equals the spontaneous emission rate ($\gamma_{ab}=\Gamma_{ba}^{sp}$), the extinction cross-section equals the scattering cross-section. In this limit, light is scattered without absorption.

\section{Field response: Green's function}
\label{sect_Green}

While the electrodynamic response of a dipole emitter (or scatterer) is described by its polarizability, the linear response of the environment is conveniently described using the electric Green's function $\bm{G}$ (also denoted by field susceptibility). The tensor (electric) Green function is defined as the solution of the vector Helmoltz equation~\cite{novotny_principles_2012,morse_methods_1953-1}
\begin{equation}
\bm{\nabla} \times \bm{\nabla} \times \bm{G}(\bm{r},\bm{r}',\omega)- \frac{ \omega^2 }{c^2} \epsilon(\bm{r},\omega) \; \bm{G}(\bm{r},\bm{r}',\omega) = \delta (\bm{r}-\bm{r}') \bm{I} \; ,
\end{equation}
satisfying the outgoing condition when $|\bm{r}-\bm{r}'| \to \infty$ (one also refers to it as the retarded Green function).
In this equation, $\delta(...)$ is the Dirac delta function, $\bm{I}$ the unit tensor and $\epsilon(\bm{r},\omega)$ is the space and frequency-dependent dielectric function of the medium. Physically, the Green function connects a monochromatic electric dipole source $\bm{d}(\omega)$ located at a position $\bm{r}'$ to the radiated electric field at a position $\bm{r}$ in the medium through the relation~\cite{carminati_electromagnetic_2015}
\begin{equation}
\bm{E}(\bm{r},\omega) = \mu_0 \omega^2  \bm{G}(\bm{r},\bm{r}',\omega) \, \bm{d}(\omega) \; .
\end{equation} 
Note that this relation holds both for classical dipoles and fields, and for quantum operators (the Green function is the same in classical and quantum electrodynamics).
The Green function contains the electrodynamic response of the environment, and can be used to relate one or several dipole sources to the electric field in arbitrary geometries such as a cavity, an antenna, an interface supporting surface plasmons or a more complex medium, that can all be treated formally on the same footing.
It will be convenient to decompose the Green function as follows:
\begin{equation}
\bm{G}(\bm{r},\bm{r}',\omega)=\bm{G}_0(\bm{r},\bm{r}',\omega)+\bm{S}(\bm{r},\bm{r}',\omega) \; ,
\end{equation}
where $\bm{G}_0$ is the free-space Green function and $\bm{S}$ is the change in the Green function due to the structured environment. Given the response of the dipole emitter (polarizability) and of the environment (Green function), we will now see that the coupling between them can be studied formally based on a picture borrowed from scattering theory \cite{foldy_scattering_1945,lax_scattering_1951}.

\section{Dipole emitter interacting with an environment}
 \label{sect_one_emitter}
 
In this section, we consider a two-level dipole emitter located at a position $\bm{r}_s$, with a fixed orientation of its transition dipole (defined by unit vector $\bm{u}$), and characterized by its free-space polarizability $\boldsymbol{\alpha}_0(\omega) = \alpha_0(\omega) \bm{u} \otimes \bm{u}$, with
\begin{equation}
\alpha_0(\omega) = \frac{3 \pi c^3}{\omega_0^3} \frac{\Gamma_0}{\omega_0-\omega-i \gamma_0/2} \; .
\label{eq_alpha_two_level}
\end{equation}
In this expression we assume $\omega \simeq \omega_0$, and we can use $\gamma_0 \geq \Gamma_0$ to account for non-radiative dephasing processes. We stress that this expression of the polarizability can also describe classical resonant scatterers \cite{de_vries_point_1998}. 

\subsection{Coupling equation}

\begin{figure}[htbp]
\centering
\includegraphics[width=8cm]{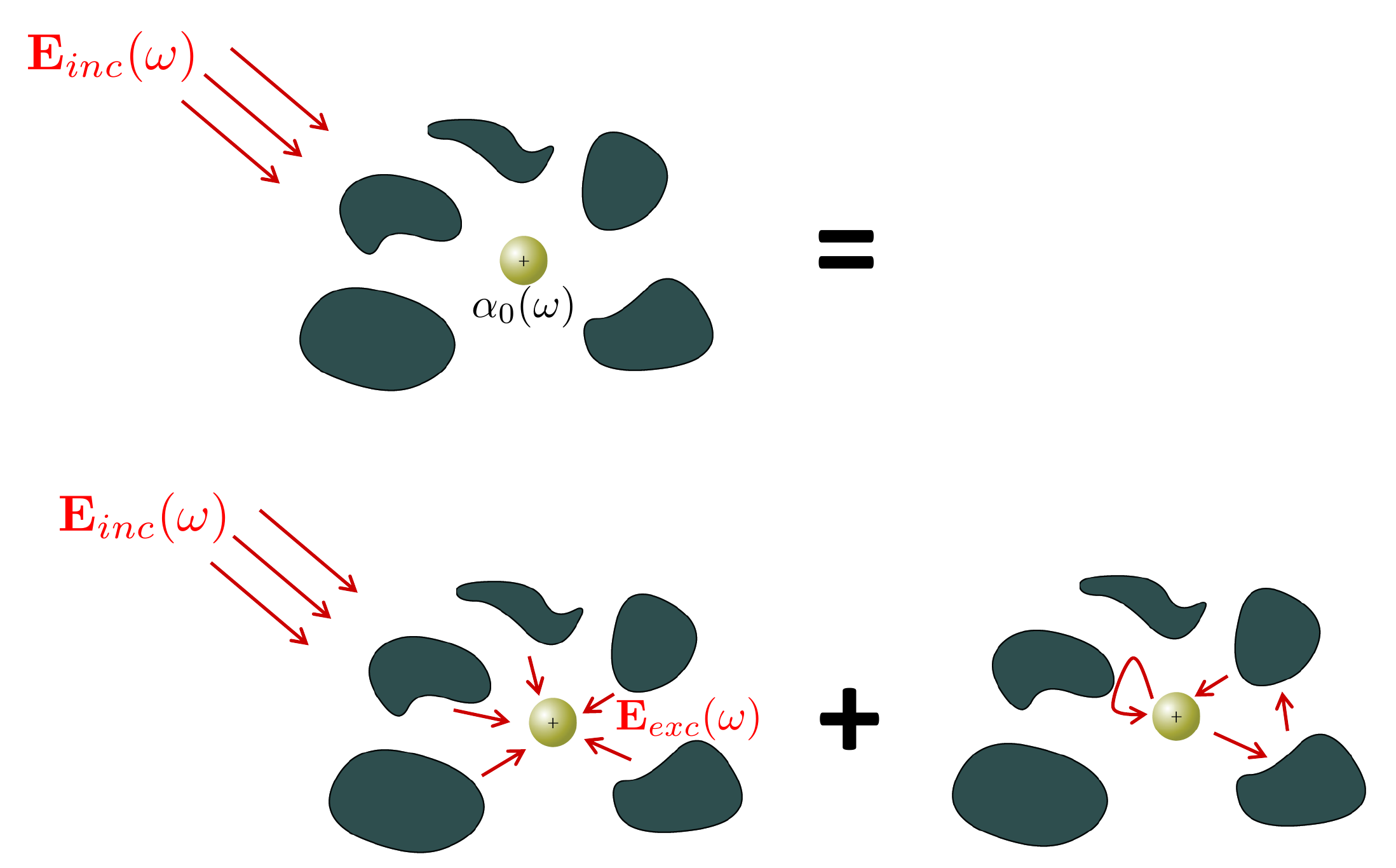}
\caption{Representation of the two scattering processes involved in the electrodynamic interaction between a dipole emitter and a structured environment.}
\label{fig_scattering}
\end{figure}

The response of the dipole emitter to an external field can be understood as a two-step scattering process. First, the emitter is excited by the field $\bm{E}_{exc}$ generated by scattering of the incident field $\bm{E}_{inc}$ by the environment. Second, the emitter is excited by its own field scattered back by the environment. These two processes are represented schematically in \fig{fig_scattering}.
With these two processes in mind, the induced dipole can be written as
\begin{equation}
\begin{split}
\bm{d}^{(+)}(\omega) =&  \boldsymbol{\alpha}_{0}(\omega) \epsilon_0 \, \bm{E}^{(+)}_{exc}(\bm{r}_s,\omega) \\
 &+ \boldsymbol{\alpha}_{0}(\omega) k_0^2 \,  \bm{S}(\bm{r}_s,\bm{r}_s,\omega) \bm{d}^{(+)}(\omega) \; ,
\end{split}
\label{eq_induced_dipole}
\end{equation}
where $k_0=\omega/c$. Note that the interaction with the environment is described by the modification of the Green function $\bm{S} = \bm{G}-\bm{G}_0$ since the interaction of the emitter with itself through the vacuum field is already included in $\boldsymbol{\alpha}_0(\omega)$. We can also define the dressed polarizability $\boldsymbol{\alpha}(\omega)$ such that
\begin{equation}
\bm{d}^{(+)}(\omega) = \boldsymbol{\alpha}(\omega) \epsilon_0 \, \bm{E}^{(+)}_{exc}(\bm{r}_s,\omega) \; .
\label{eq_dressed_polar}
\end{equation}
From Eqs.~\eqref{eq_induced_dipole} and \eqref{eq_dressed_polar}, we obtain
\begin{equation}
\boldsymbol{\alpha}(\omega)^{-1} = \boldsymbol{\alpha}_0(\omega)^{-1} - k_0^2 \,  \bm{S}(\bm{r}_s,\bm{r}_s,\omega) \; ,
\end{equation}
which gives a general expression of the dressed polarizability. Eigenmodes of the coupled system can be defined as poles in $\boldsymbol{\alpha}(\omega)$,
or zeros of $\boldsymbol{\alpha}(\omega)^{-1}$. This leads to the following general coupling equation: 
\begin{equation}
\frac{\omega^2}{c^2} \bm{S}(\bm{r}_s,\bm{r}_s,\omega)\boldsymbol{\alpha}_0(\omega) = \bm{I} \; .
\end{equation}
Projecting on the direction $\bm{u}$ of the dipole, this can be rewritten as
\begin{equation}
\frac{\omega^2}{c^2} [\bm{u} \cdot \bm{S}(\bm{r}_s,\bm{r}_s,\omega) \bm{u} ] \alpha_0(\omega) = 1 \; ,
\label{eq_coupling}
\end{equation}
which is a scalar equation. The solutions of this coupling equation, considered as an implicit equation in $\omega$, define the eigenfrequencies of the coupled system. 
For a two-level system, introducing \eqref{eq_alpha_two_level} into \eqref{eq_coupling}, we find that the complex eigenfrequencies $\varpi_p$ are solution of
\begin{equation}
\mathcal{S}(\varpi_p) = {\omega_{0} - \varpi_p - i \gamma_{0}/2} \; ,
\label{eq_coupling_two_level}
\end{equation}
where we use the notation 
\begin{equation}
\mathcal{S}(\omega)=\frac{3 \pi c \Gamma_0}{\omega_0} \left[ \bm{u} \cdot \bm{S}(\bm{r}_0,\bm{r}_0,\omega) \bm{u} \right] \; .
\end{equation}
Note that we can use $\omega=\omega_0$ in prefactors since we already assumed $\omega \simeq \omega_0$ in \eq{eq_alpha_two_level}.
Solving \eq {eq_coupling_two_level} allows one to find the complex eigenfrequencies $\varpi_p = \omega_p - i \gamma_p/2$, defining the central frequencies $\omega_p$ and the linewidth $\gamma_p$ of the eigenmodes for the coupled emitter-field system \cite{haroche_cavity_1992}.
This leads to a simple description of different interaction regimes and their main features. 

\subsection{Weak coupling}

Let us first consider the situation in which the environment has a smooth frequency dependence at the scale of the emitter linewidth $\gamma_0$. We can assume $\mathcal{S}(\omega) \simeq \mathcal{S}(\omega_0)$, and the solution to \eq{eq_coupling_two_level} simply becomes
\begin{equation}
\varpi_p = \omega_0- \frac{i}{2} \gamma_0 - \mathcal{S}(\omega_0) \; .
\label{eq_weak_single}
\end{equation}
Both the resonance frequency and the linewidth of the emitter are affected by the coupling, and are respectively given by
\begin{equation}
\omega_p = \omega_0 - \re \left[ \mathcal{S}(\omega_0) \right] \; ,
\end{equation}
\begin{equation}
\gamma_p= \gamma_0 + 2 \im \left[ \mathcal{S}(\omega_0) \right] \; .
\end{equation}
We can see that the coupling induces a (classical) frequency shift $\delta\omega = \omega_p - \omega_0$ that scales with the real part of the Green function due to the environment. The linewidth is also modified, the change scaling with the imaginary part of the Green function. In the absence of non-radiative dephasing processes $(\gamma_0 = \Gamma_0$), the change in the linewidth (or, equivalently, in the spontaneous decay rate) can be rewritten as 
\begin{equation}
\frac{\gamma_p}{\Gamma_0} =  1 + \frac{2 \im \left[ \mathcal{S}(\omega_0) \right]}{\Gamma_0}\; .
\label{eq_Gamma_weak}
\end{equation}
Introducing the partial (or projected) local density of states (LDOS), which is defined by $\rho_{\bm{u}}(\bm{r},\omega) = 2\omega/(\pi c^2) \im \left[ \bm{u} \cdot \bm{G}(\bm{r},\bm{r},\omega) \bm{u} \right]$~\cite{carminati_electromagnetic_2015}, we find
\begin{equation}
\frac{\gamma_p}{\Gamma_0} =  \frac{\rho_{\bm{u}}(\bm{r}_s,\omega)}{\rho_{\bm{u},0}(\bm{r}_s,\omega)} \; ,
\label{eq_Gamma_weak_LDOS}
\end{equation}
where $\rho_{\bm{u},0}(\bm{r}_0,\omega) = \omega^2/(3 \pi^2 c^3)$ is the partial LDOS in vacuum. We recover the well-known fact that in the weak-coupling regime, the spontaneous decay rate is modified according to the change in the LDOS, which is known as the Purcell effect (the original paper by Purcell considers the particular case of a single mode cavity with weak losses~\cite{purcell_spontaneous_1946}, the change in the LDOS being given in this case by the so-called Purcell factor). 

\subsection{Strong coupling}

We now assume that the emitter is coupled to an environment exhibiting sharp resonances, and is resonant (or quasi-resonant) with a specific mode so that we can restrict the problem to the interaction with a single mode. Assuming $\lvert \omega-\omega_m \rvert \ll \omega_m $ and $\gamma_m \ll \omega_m$, where $\omega_m$ and $\gamma_m$ are respectively the central frequency and linewidth of the mode, we can use the following single-mode expansion of the Green function
\begin{equation}
\bm{G}(\bm{r},\bm{r}',\omega)= \frac{c^2}{2 \omega_m} \, \frac{ \bm{e}_m(\bm{r}) \otimes \bm{e}_m^*(\bm{r}') }{\omega_m-\omega-i \gamma_m/2} \; ,
\label{eq_Green_mode}
\end{equation}
where $\bm{e}_m(\bm{r})$ is the normalized complex amplitude of the mode~\cite{carminati_electromagnetic_2015}. The change in the Green function $\bm{S}$ can be deduced from \eq{eq_Green_mode} by subtracting the contribution of the vacuum Green function $\bm{G}_0$ (only the imaginary part has to be subtracted, since the singular real part of $\bm{G}_0$ is not included in expression \eq{eq_Green_mode}. For a discussion of this point see \cite{wubs_multiple-scattering_2004,fussell_influence_2008}). This leads to
\begin{equation}
\mathcal{S}(\omega)=  \frac{F_m \Gamma_0 \gamma_m/4}{\omega_m-\omega-i \gamma_m/2} - i \Gamma_0 /2 \; ,
\label{eq_S_mode}
\end{equation}
where we have introduced the Purcell factor of the mode defined by~\cite{carminati_electromagnetic_2015}
\begin{equation}
F_m = \frac{6 \pi c^3}{\omega_m^2 \gamma_m} \left\lvert\bm{e}_m(\bm{r}_s) \cdot \bm{u}\right\rvert^2 \; .
\end{equation}
Note that in this definition, the factor $\left\lvert\bm{e}_m(\bm{r}_s) \cdot \bm{u}\right\rvert^2$, whose inverse defines the mode volume, depends on the emitter location and orientation (the Purcell factor is often defined using the maximum value of $\left\lvert\bm{e}_m(\bm{r}_s) \cdot \bm{u}\right\rvert^{-2}$, a convention that we do not use here). Introducing \eq{eq_S_mode} into the coupling equation \eqref{eq_coupling_two_level}, we find that the complex eigenfrequencies $\varpi_p$ of the coupled system must satisfy
\begin{equation}
\begin{split}
1=& \frac{F_m \Gamma_{0}\gamma_m /4}{ ( \omega_{0}-\varpi_p-i\gamma_{0}/2) ( \omega_{m}-\varpi_p-i\gamma_{m}/2)} \\
 - & \frac{i \Gamma_{0}/2}{  \omega_{0}-\varpi_p-i\gamma_{0}/2}  \; .
 \end{split}
\end{equation}
Solving this second-order equation, we obtain two solutions $\varpi_p^+$ and $\varpi_p^-$ given by
\begin{equation}
\begin{split}
\varpi_p^\pm =&\frac{ \varpi'_0 +\varpi_m}{2} \pm \sqrt{ \left( \frac{ \varpi_m - \varpi'_0 }{2}  \right)^2 + g^2 } \; ,
\end{split}
\label{eq_2freq}
\end{equation}
where $g = \sqrt{F_m \Gamma_0 \gamma_m/4}$ is the coupling constant, $\varpi_m = \omega_m - i \gamma_m/2$ is the complex frequency of the mode, and $\varpi'_0 = \omega_0 - i (\gamma_0 - \Gamma_0)/2$ characterizes the emitter. For $4 g^2 \ll |\varpi_m-\varpi'_0|^2$, developing the square-root term to first order, we would find two slightly modified eigenmodes (compared to the decoupled emitter and field mode), with a small frequency shift and a broadening, thus recovering the features of the weak coupling regime. In contrast, for $4g^2 \gg |\varpi_m-\varpi'_0|^2$ corresponding to the strong coupling regime, the central frequency and the linewidth of the eigenmodes become
\begin{equation}
\omega_p^\pm=  \frac{\omega_0 + \omega_m}{2} \pm \sqrt{\frac{F_m \Gamma_0 \gamma_m}{4}} \; ,
\label{eq_splitting}
\end{equation}
\begin{equation}
\gamma_p^\pm =  \frac{(\gamma_0 - \Gamma_0) + \gamma_m}{2} \; .
\label{eq_splitting_linewidth}
\end{equation}
Equation \eqref{eq_splitting} shows the appearance of two new eigenmodes of the strongly coupled system, with resonance frequencies splitted around the average resonance frequency of the uncoupled systems. Frequency splitting is a feature of the strong coupling regime, which can be experimentally observed when the splitting is larger than the linewidth of the new eigenmodes. Note that the strong coupling condition $4g^2 \gg |\varpi_m-\varpi'_0|^2$ often ensures that the frequency splitting can be experimentally observed, but is not always sufficient (for instance when $\varpi_m \sim \varpi'_0$).

\begin{figure}[htbp]
\centering
\includegraphics[scale=0.85]{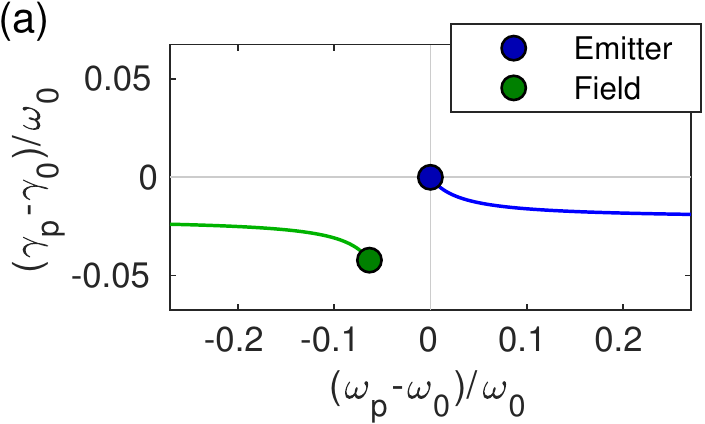}\\ \vspace{0.2cm}
\includegraphics[scale=0.85]{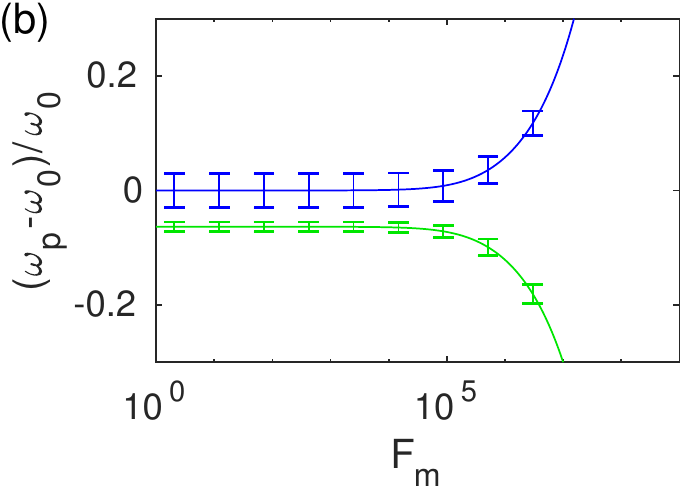}
\caption{(a)~Evolution of the eigenfrequencies of the coupled system in the complex plane when increasing the Purcell factor $F_m$ of the cavity. (b)~Normalized frequency shift of the two eigenmodes versus the Purcell factor $F_m$. Error bars represent intervals bounded by $\omega_p \pm \gamma_p/2$. }
\label{fig_strong_coupling}
\end{figure}

For the sake of illustration, let us consider a dipole emitter characterized by a central frequency $\omega_0=2370$~meV, a radiative linewidth $\Gamma_0=0.004$~meV and a total linewidth $\gamma_0=140$~meV (this values are typical of a fluorescent molecule at room temperature). We assume the emitter to be coupled to a single-mode cavity characterized by $\omega_m=2220$~meV and $\gamma_m=40$~meV. By increasing the Purcell factor $F_m$ of the cavity, we can follow the evolution of the eigenfrequencies in the complex plane, as shown in \fig{fig_strong_coupling}(a). Both the frequency splitting and the change in the linewidth are observed. The dependence of the frequency splitting on the Purcell factor (that changes the coupling constant) is shown in \fig{fig_strong_coupling}(b). In this example, the critical Purcell factor, which separates the weak and strong coupling regimes, is on the order of $10^5$ . 
%

\section{Two emitters in a structured medium}
\label{sect_two_emitters}
 
In this section we describe the interaction between two dipole emitters in a environment, and discuss strong and weak coupling regimes. In the weak coupling regime, we show that the formalism encompasses the process of irreversible energy transfer between a donor and an acceptor.

\subsection{Coupling equation}

We consider two dipole emitters located in an arbitrary medium, and excited by an external field. The emitters are characterized by their free-space polarizability
$\boldsymbol{\alpha}_i(\omega) = \alpha_i(\omega) \bm{u}_i \otimes \bm{u}_i$, the unit vector $\bm{u}_i$ defining the fixed orientation of the transition dipole, with
\begin{equation}
\alpha_{i}(\omega) = \frac{3 \pi c^3}{\omega_{0}^3} \frac{\Gamma_{i}}{\omega_{i}-\omega-i \gamma_{i}/2} \ \ \mathrm{for} \  i=1,2  \; .
\end{equation}
Since we are considering the quasi-resonant regime with $\omega \simeq \omega_1 \simeq \omega_2$, we use the average resonance frequency $\omega_0=(\omega_1+\omega_2)/2$ in all prefactors. We also introduce the following notations: 
\begin{equation}
\mathcal{S}_{ii}(\omega)=\frac{3 \pi c \Gamma_{i}}{\omega_0} \left[ \bm{u}_i \cdot \bm{S}(\bm{r}_i,\bm{r}_i,\omega) \bm{u}_i \right] \; ,
\end{equation}
\begin{equation}
\mathcal{G}_{ij}(\omega)=\frac{3 \pi c \sqrt{\Gamma_{i} \Gamma_{j}}}{\omega_{0}} \left[ \bm{u}_i \cdot \bm{G}(\bm{r}_i,\bm{r}_j,\omega) \bm{u}_j \right] \; .
\label{eq_def_Gij}
\end{equation}
While $\mathcal{S}_{ii}(\omega)$ describes the influence of each emitter on itself through the environment, $\mathcal{G}_{ij}(\omega)$ describes the interaction between them. Following the scattering picture used in section~\ref{sect_one_emitter}, the relation between the induced dipoles $\bm{d}_1=d_1 \bm{u}_1$ and $\bm{d}_2=d_2 \bm{u}_2$ in each emitter and the excitation field are conveniently expressed in a matrix form $\bm{M} \bm{X} = \bm{Y}$ where
\begin{equation}
\bm{X}=\begin{pmatrix}
d_1^{(+)}(\omega) \\
d_2^{(+)}(\omega)
\end{pmatrix} \; ,
\end{equation}
\begin{equation}
\bm{Y}=\begin{pmatrix}
\alpha_{01}(\omega) \epsilon_0 \bm{u}_1 \cdot \bm{E}_{exc}^{(+)}(\bm{r}_1,\omega)\\
\alpha_{02}(\omega) \epsilon_0 \bm{u}_2 \cdot \bm{E}_{exc}^{(+)}(\bm{r}_2,\omega) 
\end{pmatrix} \; ,
\end{equation}
\begin{equation}
\bm{M} = \begin{pmatrix}
1-\cfrac{\mathcal{S}_{11}(\omega)}{\omega_{1}-\omega-i \gamma_{1}/2} & -\cfrac{\mathcal{G}_{12}(\omega)}{\omega_{1}-\omega-i \gamma_{1}/2}  \\[7pt]
-\cfrac{\mathcal{G}_{21}(\omega)}{\omega_{2}-\omega-i \gamma_{2}/2} & 1-\cfrac{\mathcal{S}_{22}(\omega)}{\omega_{2}-\omega-i \gamma_{2}/2}
\end{pmatrix} \; .
\end{equation}
Note that reciprocity imposes that the Green function satisfies $\mathcal{G}_{12}(\omega)=\mathcal{G}_{21}(\omega)$. The eigenfrequencies of the coupled system are found by solving $\det [\bm{M}(\omega)]=0$. This leads to the following equation satisfied by the complex eigenfrequencies $\varpi_p $:
\begin{equation}
\begin{split}
0=&  1 -  \frac{\mathcal{S}_{11}(\varpi_p)}{\omega_{1} - \varpi_p - i \gamma_{1}/2}  -   \frac{\mathcal{S}_{22}(\varpi_p)}{\omega_{2} - \varpi_p - i \gamma_{2}/2} \\
& + \frac{\mathcal{S}_{11}(\varpi_p) \mathcal{S}_{22}(\varpi_p)-\mathcal{G}_{12}^2(\varpi_p)}{(\omega_{1}-\varpi_p-i \gamma_{1}/2)(\omega_{2}-\varpi_p-i \gamma_{2}/2)}  \; .
\end{split} 
\label{eq_coupling_two}
\end{equation}
This equation is a convenient starting point to discuss different interaction regimes.

\subsection{Weak coupling to the environment}

If the medium has a smooth dependence on frequency (no resonance), we can write $\mathcal{S}_{ii}(\omega) \simeq \mathcal{S}_{ii}(\omega_0) $ and $\mathcal{G}_{ii}(\omega) \simeq \mathcal{G}_{ii}(\omega_0) $. The two eigenfrequencies solutions of \eq{eq_coupling_two} are then given by
\begin{equation}
\varpi_p^\pm = \frac{\varpi_1 + \varpi_2}{2} \pm \sqrt{\left( \frac{\varpi_2-\varpi_1}{2} \right)^2 + \mathcal{G}_{12}(\omega_0)^2} \; ,
\label{eq15}
\end{equation}
where we have introduced $\varpi_i = \omega_{i} - i \gamma_{i}/2 - \mathcal{S}_{ii}(\omega_0)$, that corresponds to the eigenfrequency of each emitter considered alone in the environment (see \eq{eq_weak_single}). For strong dipole-dipole coupling between the emitters ($4 \mathcal{G}_{12}(\omega_0)^2 \gg |\varpi_2-\varpi_1|^2$) the central frequency and the linewidth of the two eigenmodes become
\begin{equation}
\begin{split}
\omega_p^\pm=&  \frac{\omega_1 + \omega_2}{2}  - \re \left[\frac{ \mathcal{S}_{11}(\omega_0) + \mathcal{S}_{22}(\omega_0)}{2}\right] \\
& \pm \re \left[ \mathcal{G}_{12}(\omega_0) \right] \; ,
\end{split}
\end{equation}
\begin{equation}
\begin{split}
\gamma_p^\pm =&  \frac{\gamma_1 + \gamma_2}{2}  + 2\im \left[\frac{ \mathcal{S}_{11}(\omega_0) + \mathcal{S}_{22}(\omega_0)}{2}\right]  \\
& \mp 2 \im \left[ \mathcal{G}_{12}(\omega_0) \right] \; .
\end{split}
\end{equation}
We observe two eigenmodes characterized by a frequency splitting that scales with $\re [ \mathcal{G}_{12}(\omega_0) ]$, {\it i.e.} with the strength of the electrodynamic coupling between the two dipoles. This is a feature of a strong coupling regime between the emitters. The linewidths show the appearance of both a broadened (or superradiant) mode and a narrowed (or subradiant) mode. 

\begin{figure}[htbp]
\centering
\includegraphics[scale=0.85]{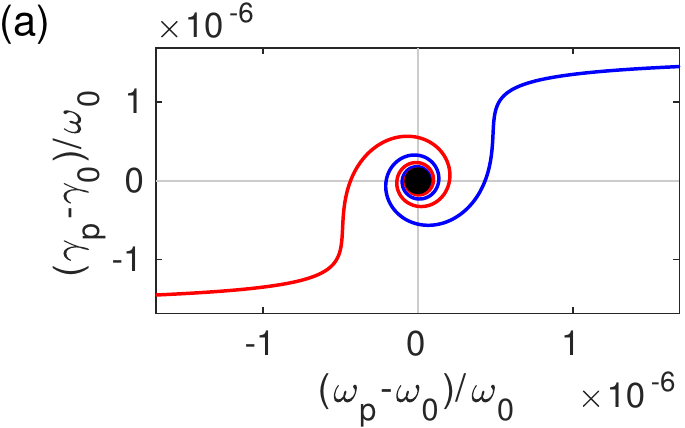}\\ \vspace{0.2cm}
\includegraphics[scale=0.85]{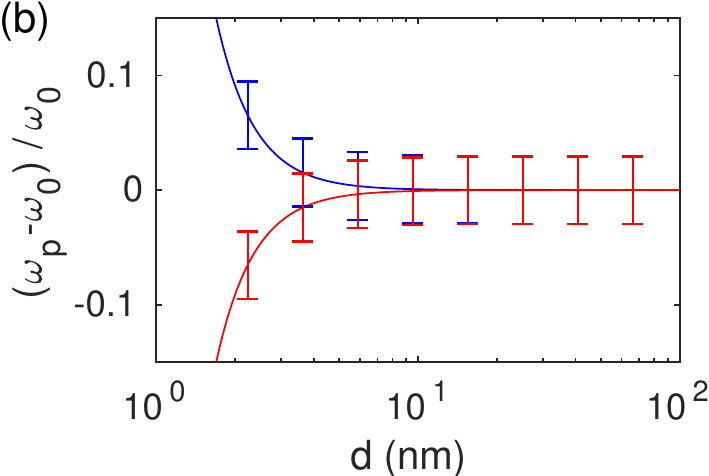}
\caption{(a)~Evolution of the eigenfrequencies in the complex plane when decreasing the distance $d$ between the emitters in free space. (b)~Normalized frequency shift of the two eigenmodes versus the distance $d$. Error bars represent intervals bounded by $\omega_p \pm \gamma_p/2$. }
\label{fig_two_emitters}
\end{figure}

To get orders of magnitude, let us take the example of two emitters in free space, with the same parameters as in the previous section, that are typical for fluorescent molecules at room temperature (central frequency $\omega_1=\omega_2=2370$~meV, radiative linewidth $\Gamma_1=\Gamma_2=0.004$~meV and total linewidth $\gamma_1=\gamma_2=140$~meV). Let us assume that the transition dipoles are oriented along the $z$-axis, and separated by a distance~$d$ along a perpendicular direction (the $x$-axis). In these conditions, the critical distance for the observation of a frequency splitting is 3~nm, and the change in the linewidth is negligible (see \fig{fig_two_emitters}). For $d \gg 3$~nm, the emitters can be considered independent. Also note that the condition $4 \mathcal{G}_{12}(\omega_0)^2 \gg |\varpi_2-\varpi_1|^2$, which we used to define the strong dipole-dipole interaction regime, is not sufficient for the observation of the frequency splitting (that has to be larger than the linewidth). 

\subsection{Weak dipole-dipole interaction}

On top of the assumption of weak coupling to the environment, we now assume that the two emitters are weakly coupled to each other ($4 \mathcal{G}_{12}(\omega_0)^2 \ll |\varpi_2-\varpi_1|^2$). In this limit, we can perform a first-order expansion of the square-root in \eq{eq15}, yielding
\begin{equation}
\varpi_p^\pm = \frac{\varpi_1 + \varpi_2 \pm (\varpi_2-\varpi_1)}{2} \pm  \frac{ \mathcal{G}_{12}(\omega_0)^2}{\varpi_2-\varpi_1} \; .
\end{equation}
We see that the eigenmode $+$ (resp. $-$) correponds to the modifications in frequency and linewidth of emitter $2$ (resp. $1$).
It is interesting to show that this expression can describe irreversible energy transfer between two emitters (usually referred to as donor and acceptor), that at short distance is known as F\"orster resonant energy transfer (FRET)~\cite{forster_zwischenmolekulare_1948}. FRET has been widely used in biology as a mechanism to detect molecular interaction~\cite{selvin_renaissance_2000}. Short-distance energy transfer is also involved in the process of photosynthesis~\cite{scholes_lessons_2011}. To compute the eigenfrequencies in this regime, we need to specify the polarizability models for emitter~1 (donor) and emitter~2 (acceptor). We shall assume that the donor is an ideal two-level atom (with $\gamma_1=\Gamma_1$), while the acceptor is a three level system, with a large excited-state decay rate towards the auxiliary radiative level (as in a florescent molecule). This means that the condition 
$\gamma_2 \gg \left( \Gamma_{1}, \Gamma_{2},2 \im \left[ \mathcal{S}_{11}(\omega_0) \right],2 \im \left[ \mathcal{S}_{22}(\omega_0) \right] \right)$ is assumed to be satisfied. Note that, as described in section~\ref{sect_polar}, the polarizability $\alpha_2(\omega)$ describes the excitation of emitter~2 only (subsequent fluorescent emission at a different frequency is implicit). We also assume $\omega_1=\omega_2=\omega_0$, meaning that the emission frequency of the donor matches the absorption frequency of the acceptor. Under these conditions, the linewidth of the eigenmodes of the coupled system 	are 
\begin{equation}
\gamma_p^+ = \gamma_2 \; ,
\end{equation}
\begin{equation}
\gamma_p^- = \gamma_1 + 2 \im \left[ \mathcal{S}_{11}(\omega_0) \right] + \frac{4 \re \left[ \mathcal{G}_{12}(\omega_0)^2 \right] }{\gamma_2} \; .
\label{eq_linewidth_FRET}
\end{equation}
As expected, eigenmode $+$ (corresponding to emitter~2 or acceptor) has a negligible broadening due the coupling to both the donor and the environment (this follows directly from the condition of a large $\gamma_2$). More interestingly, the linewidth associated to eigenmode $-$ (corresponding to emitter~1 or donor) is modified by the surrounding medium (second term in the right-hand side in \eq{eq_linewidth_FRET}) and by the presence of the acceptor (third term in the right-hand side in \eq{eq_linewidth_FRET} which will be denoted by $\Gamma_{inter}$). We can observe that the presence of the acceptor can either increase or decrease the linewidth of the donor, depending on the sign of $\re \left[ \mathcal{G}_{12}(\omega_0)^2 \right]$. This can be understood as the result of changes in the relative phase between the induced dipole in the donor and the field backscattered by the acceptor at the donor position, as in the process giving rise to oscillations in the fluorescence lifetime of an emitter in front of a reflective interface~\cite{drexhage_influence_1970}. For distances much smaller than the wavelength $\lambda_0=2\pi c/\omega_0$, we can assume $\re [ \mathcal{G}_{12}(\omega_0)^2 ] \simeq \left \vert \mathcal{G}_{12}(\omega_0) \right \vert ^2$ (this can be easily verified in free space, as long as the interdistance $d < \lambda_0/4$). This means that at short distance the linewidth of the donor is always increased by the presence of the acceptor. Moreover, from Eqs.~\eqref{eq_extinction} and~\eqref{eq_scattering}, one can deduce the on-resonance expressions of the extinction and scattering cross-sections $\sigma_{e}(\omega_0) $ and $\sigma_{s}(\omega_0) $, which are respectively
\begin{equation}
\sigma_{e}(\omega_0) = \frac{6 \pi c^2}{\omega_0^2 }  \frac{\Gamma_2}{\gamma_2}\; ,
\end{equation}
\begin{equation}
\sigma_{s}(\omega_0) = \frac{6 \pi c^2}{\omega_0^2 }\left(\frac{\Gamma_2}{\gamma_2} \right)^2 \; .
\end{equation}
In the regime $\gamma_2 \gg \Gamma_2$, the scattering cross-section is negligible, and we can assume that the absorption cross-section $\sigma_{a}(\omega_0)$ equals the extinction cross-section. Then, the last term in the right-hand side in \eq{eq_linewidth_FRET}, usually referred to as the energy transfer rate $\Gamma_{et}$, can be written
\begin{equation}
\Gamma_{et}= 6\pi \, \Gamma_1 \sigma_{a}(\omega_0) \left | \bm{u}_1 \cdot \bm{G}(\bm{r}_1,\bm{r}_2,\omega_0) \bm{u}_2 \right |^2  \; ,
\label{eq_fret_rate}
\end{equation}
where we have used \eq{eq_def_Gij}.
This expression takes the usual form of the energy transfer rate in dipole-dipole interaction~\cite{novotny_dipole-dipole_2012} (see also \cite{dung_resonant_2002,dung_intermolecular_2002} for a full QED treatment). The main difference between $\Gamma_{et}$ and $\Gamma_{inter}$ is that the latter includes a back-action from the acceptor to the donor due to scattering, that disappears in the energy transfer regime.
 In free space, the Green function at short distance can be taken in the quasi-static limit, and follows the scaling $\bm{u}_1 \cdot \bm{G}_0(\bm{r}_1,\bm{r}_2,\omega_0) \bm{u}_2 \sim |\bm{r}_1-\bm{r}_2|^3$. The free-space energy transfer rate therefore scales as $\Gamma_{et} \sim |\bm{r}_1-\bm{r}_2|^6$, which is a feature of FRET, as initially derived by F\"orster~\cite{forster_zwischenmolekulare_1948}. In more complex geometry, inserting the appropriate Green function into \eq{eq_fret_rate} allows one to compute the change in the FRET rate due to the environment (see for example \cite{bouchet_long-range_2016,vincent_magneto-optical_2011}).

\begin{figure}[htbp]
\centering
\includegraphics[scale=0.85]{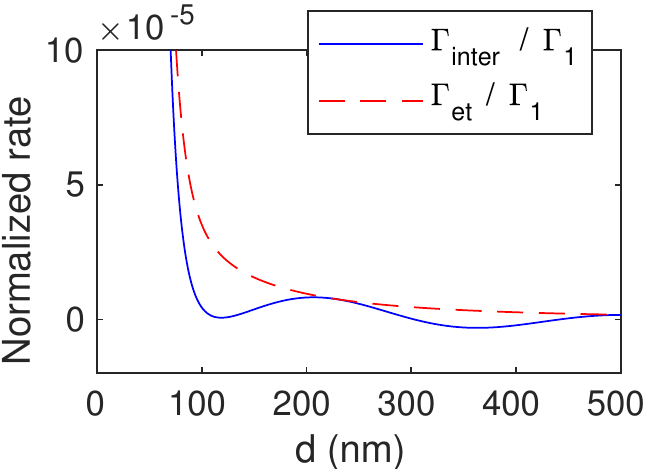}
\caption{Solide line: Normalized modification in the donor linewidth due to the presence of the acceptor $\Gamma_{inter}/\Gamma_1$ versus the distance $d$ between donor and acceptor. Dashed line: Normalized energy transfer rate $\Gamma_{et}/\Gamma_1$ calculated using expression \eqref{eq_fret_rate}.}
\label{fig_FRET}
\end{figure}

As a didactic example, let us consider a donor (two-level system) with emission frequency $\omega_1=2370$~meV and radiative linewidth $\gamma_1=\Gamma_1=0.004$~meV, and an acceptor (three-level molecule) with absorption frequency $\omega_2=\omega_1$ and total linewidth $\gamma_2=140$~meV.
We show in \fig{fig_FRET} the energy transfer rate $\Gamma_{et}$, calculated using \eq{eq_fret_rate}, versus the distance $d$ between donor and acceptor. For comparison, we also display the change in the donor linewidth due to the acceptor that includes the scattering back-action ($\Gamma_{inter}$). We see that both expressions coincide for $d<100$~nm. Note that for larger distances, the difference would remain difficult to observe since the energy transfer efficiency is very low in this regime (on the order of $10^{-6}$).

\subsection{Strong dipole-field interaction}

\begin{figure}[htbp]
\centering
\includegraphics[scale=0.85]{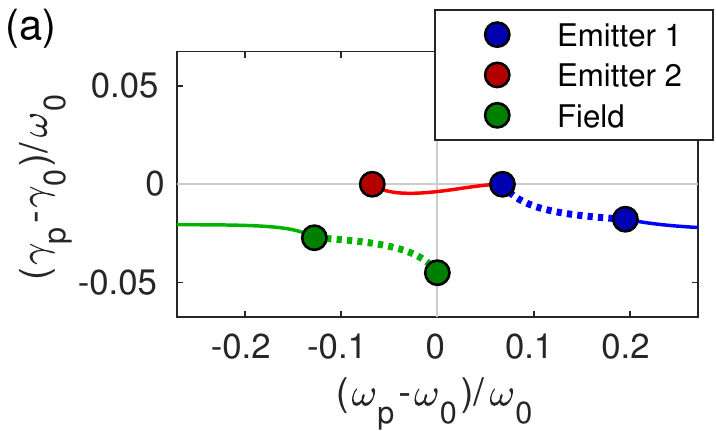}\\ \vspace{0.2cm}
\includegraphics[scale=0.85]{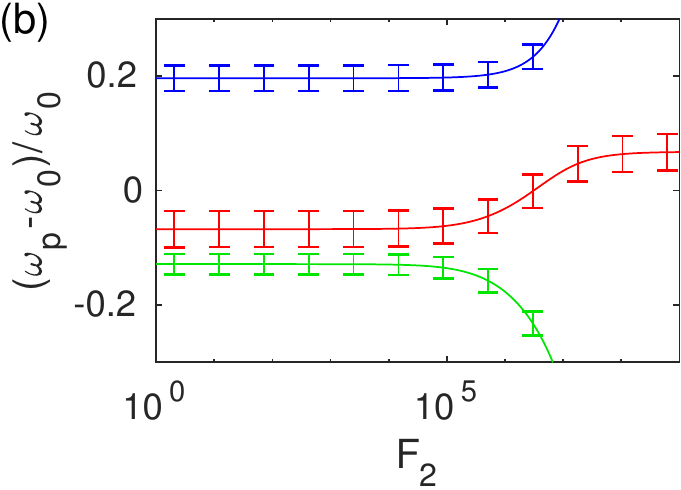}
\caption{(a)~Evolution of the eigenfrequencies in the complex plane when increasing the Purcell factor $F_2$ of emitter~2. (b)~Normalized frequency shift of the three eigenmodes versus the Purcell factor $F_2$. Error bars represent intervals bounded by $\omega_p \pm \gamma_p/2$. }
\label{fig_strong_coupling_two}
\end{figure}

We now examine the regime of strong coupling of the two emitters to a single electromagnetic mode. To proceed, we use the expansion of the Green function in \eq{eq_Green_mode}. For convenience we introduce the Purcell factor experienced by each emitter, defined by
\begin{equation}
F_i = \frac{6 \pi c^3}{\omega_m^2 \gamma_m} \left\lvert\bm{e}_m(\bm{r}_i) \cdot \bm{u}_i\right\rvert^2 \ \ \mathrm{for} \  i=1,2 \; .
\end{equation}
It follows that
\begin{equation}
\mathcal{S}_{ii}(\omega)=  \frac{F_i \Gamma_i \gamma_m /4}{\omega_m-\omega-i \gamma_m/2} - i \Gamma_i /2 \; ,
\end{equation}
\begin{equation}
\mathcal{G}_{ij}(\omega)^2=\frac{F_i \Gamma_i F_j \Gamma_j \gamma_m^2/16}{(\omega_m-\omega-i \gamma_m/2)^2} \; .
\end{equation}
The eigenfrequencies of the coupled system can then be found by inserting these expressions into \eq{eq_coupling_two}, resulting in a third-order equation in the complex frequency $\varpi_p$ whose roots can be found analytically. Different behaviors can be observed depending on the relative values of the two coupling constant $g_1$ and $g_2$, defined as
\begin{equation}
g_i = \sqrt{ \frac{F_i \Gamma_i \gamma_m}{4}} \ \ \mathrm{for} \  i=1,2 \; .
\end{equation}
Indeed, if $g_1$ and $g_2$ are substantially different, the emitter with the larger coupling dictates the features of two splitted eigenmodes (resulting from strong coupling between this emitter and the field mode), while the third eigenmode is associated to the other uncoupled emitter. The situation is more complicated when $g_1 \sim g_2$ since in this case the features of the three eigenmodes depend on both emitters. As an example, let us consider two emitters with features matching those of fluorescent molecules at room temperature, characterized by different resonant frequencies ($\omega_1=2370$~meV and $\omega_2=2070$~meV), by a radiative linewidth $\Gamma_1=\Gamma_2=0.004$~meV and by a total linewidth $\gamma_1=\gamma_2=140$~meV. We assume the emitters coupled to a single-mode cavity with $\omega_m=2220$~meV and $\gamma_m=40$~meV. Moreover, we set the Purcell factor seen by emitter~1 to $F_1=3 \times 10^6$ so that this emitter is strongly coupled to the mode (see \fig{fig_strong_coupling}), while $F_2$ is left as a free parameter allowing us to tune the coupling strength of emitter~2. The behavior of the central frequency $\omega_p$ and linewidth $\gamma_p$ of the three eigenmodes is shown in \fig{fig_strong_coupling_two}.
For $F_2 \ll F_1$, we observe the two eigenfrequencies resulting from the strong coupling between the emitter~1 and the field mode, while the third eigenfrequency is associated with the unperturbed emitter~2. By increasing $F_2$, this third eigenfrequency progressively changes to the free-space resonance frequency of emitter~1. In the regime $F_2 \gg F_1$, two eigenfrequencies characterizes strong coupling between emitter~2 and the field emerge, while the third eigenfrequency is associated to the unperturbed emitter~2. This behavior can be interpreted as follows: The emitter with the largest coupling constant strongly couples to the field mode, creating two frequency shifted new eigenmodes, leaving the other emitter out of resonance.
%

\section{Generalization: $N$ identical dipole emitters in mutual interaction}
\label{sect_many_emitters}
 
The approach can be extended to $N$ dipole emitters coupled to a structured environment. In the simplest situation, we can assume $N$ identical emitters with a polarizability
$\boldsymbol{\alpha}_0(\omega)$ given by \eq{eq_alpha_two_level}, all with the same orientation of their transition dipole. We can also assume that all emitters see the same environment, so that $\mathcal{S}_{ii}(\omega)=\mathcal{S}(\omega)$ and $\mathcal{G}_{ij}(\omega)=\mathcal{G}(\omega)$. Finding the eigenfrequencies of the coupled system amounts to solve $\det (\bm{M})=0$, where $\bm{M}$ is now a $N \times N$ matrix. This leads to the following equation for the complex eigenfrequencies $\varpi_p$:
\begin{align}
\left( 1 - \frac{\mathcal{S}(\varpi_p)-\mathcal{G}(\varpi_p)}{\omega_0-\varpi_p-i \gamma_0/2} \right)^{N-1} & \\
\times \left( 1 -  \frac{\mathcal{S}(\varpi_p)+(N-1) \mathcal{G}(\varpi_p)}{\omega_0-\varpi_p-i \gamma_0/2} \right) &= 0 \; .
\end{align}
Whenever the surrounding medium can be considered as weakly resonant, the $N$ eigenfrequencies are given by 
\begin{equation}
\varpi_p^-= \omega_0 - \frac{i}{2}\gamma_0-  \mathcal{S}(\omega_0) - (N-1)  \mathcal{G}(\omega_0)  \; ,
\end{equation}
\begin{equation}
\varpi_p^+=\omega_0 - \frac{i}{2}\gamma_0-  \mathcal{S}(\omega_0)+   \mathcal{G}(\omega_0) \; ,
\label{eq2}
\end{equation}
where the the solution $\varpi_p^+$ has a multipicity $N-1$. In contrast, if the surrounding medium is strongly resonant, and the emitters are quasi-resonant with one specific eigenmode $m$ of the field, we obtain a very different collective behavior. Using the notations $\varpi'_0 = \omega_0 -i (\gamma_0-\Gamma_0)/2$ and $\varpi_m = \omega_m -i \gamma_m/2$, and introducing the coupling constant $g=\sqrt{\Gamma_0 \gamma_m F_m/4}$ with $F_m$ the Purcell factor of the mode, we find $N+1$ eigenfrequencies given by
\begin{equation}
\varpi_p^{\pm}=\frac{\varpi'_0+\varpi_m}{2}  \pm  \sqrt{\left( \frac{\varpi_m-\varpi'_0}{2} \right)^2 + N g^2  } \; ,
\label{eq_Nfreq}
\end{equation}
\begin{equation}
\varpi_p'=\varpi'_0 \; ,
\label{eq_N}
\end{equation}
where the solution $\varpi_p'$ has a multiplicity $N-1$. The eigenfrequencies given by \eq{eq_Nfreq} can be compared to the solutions for one emitter strongly coupled to the field given by \eq{eq_2freq}. These solutions only differ by a modification of the coupling constant, and the coupling constant for $N$ identical emitters is simply $\sqrt{N} g$, where $g$ is the coupling constant for one emitter.
Interestingly, the weak-coupling and strong-coupling situations lead to very different results. In particular, when the environment is weakly resonant, the splitting in frequency and the linewidth scale with $N$, as one would obtain with the theory of Dicke superradiance in the weak-excitation limit~\cite{gross_superradiance:_1982}. In contrast, when the environment is strongly resonant, the splitting scales with $\sqrt{N}$ and the linewidth does not depend on $N$, in agreement with results obtained with the Jaynes-Cumming Hamiltonian~\cite{agarwal_vacuum-field_1984}. The simple coupled-dipoles model introduced in this tutorial therefore contains the main ingredients to describe collective interactions between quantum emitters under external excitation.

\section{Conclusion}
\label{sect_conclusion}

In summary, we have presented a semi-classical description of the interaction between one or several quantum dipole emitters and a structured environment under weak external excitation. The approach is based on a self-consistent coupling equation resulting from a scattering picture. This coupling equation serves as a starting point to discuss many interaction regimes, covering weak and strong-coupling between a single emitter and the electromagnetic field, collective interactions between several emitters leading for example to superradiance, as well as energy transfer between two emitters. This simple approach provides both a unified description and an intuitive understanding of the behavior of dipole emitters in (nano)structured environments. It can also serve as a foundation for more elaborate models, including an explicit quantization of the electromagnetic field and/or saturation effects in the emitter dynamics~\cite{agarwal_quantum_1974,mandel_optical_1995,berman_QED_1994,haroche_exploring_2013}.

\section*{Acknowledgments}

This work was supported by LABEX WIFI (Laboratory of Excellence within the French Program ``Investments for the
Future'') under references ANR-10-LABX-24 and ANR-10-IDEX-0001-02 PSL*. 
DB acknowledges ESPCI Paris for a three-months postdoctoral grant that permitted to achieve this work.


\end{document}